\documentclass[10pt,twocolumn]{article}

\usepackage[numbers,sort&compress]{natbib}

\usepackage{times}
\usepackage{graphicx}
\usepackage{amsmath}
\usepackage{booktabs}
\usepackage{url}
\usepackage{tcolorbox}

\usepackage[
    top=0.75in,
    bottom=1in,
    left=0.75in,  
    right=0.75in,
    columnsep=0.25in
]{geometry}

\usepackage{titlesec}
\titleformat{\section}{\large\bfseries}{\thesection}{1em}{}
\titleformat{\subsection}{\normalsize\bfseries}{\thesubsection}{1em}{}

\title{The Ethics Engine: A Modular Pipeline for Accessible Psychometric Assessment of Large Language Models}
\author{Jake Van Clief, Constantine Kyritsopoulos}

\begin{document}
\maketitle

\begin{abstract}
As Large Language Models increasingly mediate human communication and decision-making, understanding their value expression becomes critical for research across disciplines. To help with this, I present the Ethics Engine, a modular Python pipeline that transforms psychometric assessment of LLMs from a technically complex endeavour into an accessible research tool. In this work I attempt to demonstrate how thoughtful infrastructure design can expand who participates in AI research, enabling investigators across cognitive science, political psychology, education, and other fields to study value expression in language models. Recent adoption of this tool by the Neuropolitics Lab researchers at the University of Edinburgh studying authoritarianism validates its research utility. The study processed over 10,000 AI responses to assessment questions across multiple models and contexts, investigations that would have been prohibitively complex without accessible infrastructure. I argue that such tools fundamentally change the landscape of AI research by lowering technical barriers while maintaining scientific rigor. As LLMs increasingly serve as cognitive infrastructure—tutoring children, drafting policy documents, mediating workplace communication—their embedded values shape millions of daily interactions. Without systematic measurement of these value expressions, we deploy systems whose moral influence on users, particularly developing minds, remains entirely uncharted. The Ethics Engine enables the rigorous assessment necessary for informed governance of these increasingly influential technologies.
\end{abstract}

\section{Introduction}

Machines now finish our sentences before we know what we meant to say. From classroom chatbots to courtroom briefs, large language models—GPT-4, Claude, and their kin—sit between intent and outcome, quietly nudging tone and content. I want to know what moral cargo those nudges carry. What the "ethics board" behind each of these companies decided their definition of ethics was. The Ethics Engine is my answer: a Python pipeline that questions models the way psychologists question people, then maps the answers onto familiar moral terrain.

I have watched chatbots console veterans, draft data base code, and tutor children. Each exchange is a tiny governance event: a user asks, the model answers, power flows. My proposal is therefore both technical and existential. I want to know which values ride inside the process, and where those values are decided, before society stakes more of its moral credit on invisible code.

The moral meaning behind words and the arguments about those meanings have always raked my mind, and only grew more important after engaging with Alasdair MacIntyre's critique of modern moral discourse on virtue:

\begin{quote}
\centering
\textit{You cannot hope to reinvent morality on the scale of a whole nation... when the very idiom of morality which you seek to re-invent is alien in one way to the vast mass of ordinary people and in another to the intellectual elite.}

— Alasdair MacIntyre, \textit{After Virtue}, 3rd ed., p. 238
\end{quote}

The very words we use to name the 'good' mean different things to different people. Courage against tyranny means storming the halls of a corrupt Congress on January 6th or stopping masked ICE agents from abducting people off the street, depending on who you ask. These kinds of individual differences in perceptions, attitudes, thoughts, beliefs, and values are the province of psychologists, and measuring how they differ, what these words reflect in the underlying psychology, is the purpose of psychometrics. My thesis leverages the tools of psychometrics to measure the values hidden inside LLMs. If decades of validated instruments can map the moral terrain of human populations, revealing how authoritarianism correlates with voting patterns or how moral foundations predict policy preferences, then these same instruments can illuminate the value structures encoded in the models increasingly mediating our discourse.

Psychometric assessment typically assumes relatively stable traits—asking 'how authoritarian are you?' presumes a consistent self to measure. But LLMs display remarkable context sensitivity: their responses shift dramatically based on framing, instructions, and implied audience. This sensitivity initially appears to undermine measurement entirely—if a model gives different answers depending on how you ask, what are we measuring?

This variability, however, mirrors a phenomenon psychologists have long studied in humans: demand characteristics and context-dependent self-presentation. People respond differently to personality assessments when applying for jobs versus research participation, when they believe responses are anonymous versus identified. Psychometric instruments handle this through repeated measurement across diverse question phrasings, statistically inferring stable traits from variable responses. The Ethics Engine applies this same principle to LLMs, but adds a critical dimension: systematic manipulation of the social context in which models respond—what we term 'personas.'

A persona consists of instructions that frame how the model should position itself ideologically when responding to assessment items. Rather than asking 'neutral' GPT-4 about authority (which might yield corporate-approved moderate responses), the pipeline probes: How does this model express values when instructed to respond as a progressive activist? As a libertarian technologist? As a social conservative? By varying these contextual frames while holding the underlying model constant, we map not just what values the model 'has' but how it understands value expression across different social positions—revealing what the system has learned about how different groups think, what they value, and how they communicate those values, accurate to human data or not

To me, the most important aspect of the data this pipeline generates is not the individual scores on scales or even a model's overall profile, it's the difference between extremes across models and prompts. Looking at one score or another individual output can give you conclusions that could be explained away by the inherent generative and predictive processes of LLMs. Any single response might simply reflect non-signifiant statistical patterns in training data.

By examining the extremes, how dramatically scores shift between personas, how differently models respond to identical prompts, the range between minimum and maximum expressions—reveals something deeper. These variations tell a story beyond statistical generation. When one model swings from a 2 to a 7 on authoritarianism while another stays locked at 4, when liberal personas produce minor shifts but conservative ones trigger dramatic changes, when certain value dimensions show vast inter-model variance while others remain stable, these patterns suggest systematic differences in how values are encoded and expressed. The pipeline's ability to capture these extremes at scale transforms anecdotal observations about model behaviour into mappable territories of variation that researchers can explore systematically.

This multi-context approach addresses a fundamental limitation in previous LLM assessment work. Studies administering personality inventories to 'base' ChatGPT or Claude typically report single aggregate scores—GPT-4 is ENFJ, Claude is INTJ—as if these models have unitary personalities. But such scores collapse crucial variance. When Serapio-García et al. (2023) found high variability in LLM personality responses depending on prompt phrasing, this wasn't measurement error to be minimized—it revealed that models encode multiple, context-dependent response patterns rather than singular stable traits.

The Ethics Engine treats this variability as signal rather than noise. By systematically varying both assessment items (through validated scales with multiple phrasings of similar constructs) and ideological framing (through persona prompts), the pipeline captures the full distribution of value expressions a model can produce. The range between minimum and maximum scores, the differential response to progressive versus conservative framing, the consistency of patterns across multiple runs—these distributional properties reveal how values are encoded in the model in ways single scores cannot capture.

Over the rest of this thesis I will share the how and why behind my modular, scalable implementation of an LLM Survey Aggregator(The Ethics Engine) that transforms abstract concerns about AI bias into concrete, measurable phenomena. The work addresses four primary research questions:

\begin{enumerate}
\item How can a modular, user-friendly tool reliably measure moral and ideological biases in LLMs?
\item In what ways do different prompting methods impact the expression of biases in these models?
\item How can I clearly communicate these biases to non-technical audiences—particularly policymakers and ethicists—to facilitate responsible governance?
\item What methodological challenges might emerge, and how will future research address them?
\end{enumerate}

The Ethics Engine serves two publics: engineers/researchers who tweak weights of the models and regulators who set rules. By giving both groups a common diagnostic, I aim to replace guesswork with evidence, letting us contest AI values in daylight rather than chase them in hindsight.

\section{Literature Review and Theoretical Framework}

\subsection{The Psychometric Turn in AI Assessment}

The application of psychological instruments to artificial systems represents more than methodological convenience, it acknowledges that LLMs, trained on human-generated text, inevitably absorb and reflect human psychological patterns. Recent scholarship has demonstrated that LLMs exhibit stable personality-like structures when tested using classical psychological instruments (Pellert et al., 2024; Salecha et al., 2024). These studies reveal consistent profiles marked by high Openness and Agreeableness, with relatively low Neuroticism across multiple model architectures (de Winter et al., 2024).

Critically, this correspondence goes deeper than surface-level mimicry. Binz and Schulz (2023) demonstrated in \textit{Nature Human Behaviour} that LLMs trained on appropriate human data not only replicate human responses but generate internal representations that partially correspond to human neural representations when performing similar cognitive tasks. This suggests that LLMs aren't merely outputting statistically likely responses, they appear to be approximating, to some degree, the actual cognitive processes humans use to generate those responses in relation to the meaning of the words. This deeper relative relation between human and artificial cognition of natural language provides the strongest scientific and philosophical justification for psychometric assessment of AI systems. If these models are genuinely developing human-like cognitive structures around the meaning of words through training, then the decades of validated instruments designed to measure human psychology using words become not just convenient but essential tools for understanding what we've built. The validity of the Ethics Engine, and indeed any psychometric approach to AI assessment, fundamentally rests on this correspondence between artificial and human psychological processing surrounding language. 

In no way am I making the assertion that current LLMs or any other NLP/Machine Learning systems do and will function as the human brain in terms of reasoning or thinking. Rather, I'm observing that the statistical outputs of both systems converge on remarkably similar patterns. When humans and LLMs answer personality questionnaires or moral dilemmas, they produce response distributions that correlate strongly, not because silicon mimics neurons necessarily, but because both systems have learned to navigate the same linguistic and conceptual space. A bird and an air-plane use entirely different mechanisms for flight, yet both must obey aerodynamics to stay aloft. Similarly, transformers computing attention weights and humans processing meaning through neural networks arrive at comparable outputs through allegedly different architectures. 

The mathematical signature of their responses, the means, variances, and correlations across psychological dimensions, align closely enough that validated psychometric instruments designed for one can meaningfully measure the other. This statistical correspondence, not architectural similarity, justifies applying human psychological tools to artificial systems. We're not measuring how these systems think (or if they think), but rather mapping the values embedded in what they produce. 

Further, the choice of psychometric assessment is not arbitrary, by doing so we can leverage decades of cognitive science research demonstrating that moral values follow predictable patterns across populations (Haidt, 2001; Graham et al., 2013). Importantly, moral foundations theory has been validated across diverse cultural contexts, though the expression and prioritization of these foundations varies significantly between cultures (Graham et al., 2011; Haidt et al.,1993; Iyer et al., 2012; Davies et al., 2014; Doğruyol et al., 2019). Recent work has further refined these instruments, suggesting adjustments to scoring methods and revealing additional moral dimensions beyond the original five foundations (Atari et al., 2023; Hopp et al., 2021; Curry et al., 2019).

While this work focuses on authoritarianism measures given their robust psychometric properties and clear implications for democratic governance, the modular architecture of the Ethics Engine readily accommodates other validated instruments. Future research should extend this approach to moral foundations, personality inventories, and domain-specific value assessments, building a comprehensive profile of how different models encode and express psychological constructs.

When we apply these instruments to LLMs, we're not imposing a Western-centric framework but employing tools that, despite their limitations, have demonstrated validity across human societies. The variations in how different cultures weight moral foundations, how collectivist societies prioritize loyalty and authority differently than individualist ones, provide crucial context for interpreting LLM responses. These models, trained on global internet text, likely encode multiple cultural perspectives on morality simultaneously. By applying these validated instruments to LLMs, we can map how their responses align with known structures of human moral cognition across cultures. This matters because LLMs increasingly serve as cognitive prostheses, extending human judgment in domains from medical diagnosis to legal analysis for users worldwide.

What makes these all of the above findings particularly relevant for research is their dual stability. First, different models exhibit distinct and consistent psychological profiles—ChatGPT \textit{(gpt4), }Claude\textit{(sonnet 3.5)}, and Gemini\textit{(1.5)} each express characteristic patterns that remain stable across repeated testing (Abdurahman et al., 2024; Rutinowski et al., 2024). A model that scores high on openness or low on authoritarianism tends to maintain those traits reliably, suggesting these aren't random outputs but systematic expressions embedded in the model's parameters. Second, and equally important, these profiles differ meaningfully between models, reflecting the unique combinations of training data, design choices, and potentially the political and economic priorities of their creators (Feng et al., 2023; Rozado, 2023).

This dual stability, or better put '\textit{consistency }' within models and meaningful variation between them transforms value expression from noise into signal. We're not measuring random fluctuations but systematic differences that likely trace back to decisions made in training rooms and boardrooms or even represents parts of the training data as a semantic sum. When OpenAI's models consistently express different moral foundations than Anthropic's, or when open-source models diverge from their commercial counterparts, we're glimpsing how human choices about data curation, reward functions, and safety filters crystallize into measurable psychological signatures. These stable, distinctive patterns make LLMs amenable to rigorous scientific study, precisely what the Ethics Engine enables. Rather than anecdotal observations about bias, we can now map the moral topology of different AI systems with the same precision psychologists use to study human populations.

\subsection{Prompt Sensitivity and the Methodological Challenge}

Critically, LLM outputs prove highly sensitive to subtle variations in prompts. Seemingly negligible re-phrasings significantly shift measured responses, leading to quite different inferences regarding any underlying psychometric representation(Serapio Garcia 2023). This sensitivity reveals a fundamental challenge: comprehensive assessment requires testing multiple phrasings, personas, and contexts.

This variability between phrasings might suggest LLMs simply become whoever you ask them to be—infinitely malleable systems with no stable core. But as seen above that's not what the data shows. While LLMs display far greater variability in their expressions than any human would, a person doesn't swing from extreme liberal to conservative based on word choice, they maintain statistically detectable consistencies across permutations of phrasings and inputs (Pellert et al., 2024; Salecha et al., 2024). An LLM asked about authority in ten different ways might give ten different responses, but the mean and variance of those responses reveals an underlying tendency—a psychological center of gravity that psychometric instruments can detect. 

This sensitivity to prompt framing parallels a phenomenon psychologists have extensively studied in human participants: demand characteristics. People modify their responses based on perceived social expectations—presenting differently in job interviews versus anonymous research studies, adjusting self-descriptions based on who's asking and why. A participant who scores moderate on authoritarianism in neutral conditions might express stronger authoritarian views when they believe they're among like-minded individuals, or suppress such views when anticipating social judgment.

Psychometric instruments handle this context-dependency through specific design features: administering multiple items that probe the same construct with varied wording, including reverse-scored items to detect acquiescence bias, and measuring across diverse hypothetical scenarios to statistically infer the common variance—the underlying trait—that persists across contexts.

The same participant might describe themselves as 'preferring strong leadership' in one item and 'valuing questioning authority' in another; aggregating across such items reveals their actual authoritarian tendency beneath context-specific presentation management.

LLMs exhibit an analogous but more extreme version of this phenomenon. Where humans show modest context-effects, models display dramatic shifts—the same underlying system producing responses that span from libertarian to authoritarian depending on perceived user expectations. This heightened sensitivity doesn't invalidate measurement; it necessitates methodological rigor. By systematically varying prompt contexts (personas) while administering validated multi-item scales, the Ethics Engine applies the same statistical logic psychologists use for human assessment: inferring stable patterns from variable responses, distinguishing genuine flexibility from measurement error, and mapping the boundaries of how models understand and express values across social contexts.

\subsection{From Technical Metrics to Research Questions}

Engineers have developed sophisticated metrics for evaluating language models. Perplexity scores measure how surprised a model is by test data—lower scores indicating better prediction of the next word (Hendrycks et al. 2021). Benchmark suites like MMLU test factual knowledge across domains, while HellaSwag evaluates common sense reasoning. TruthfulQA measures whether models generate accurate rather than merely plausible-sounding text. These technical achievements are impressive: they've given us models that can pass bar exams, solve mathematical proofs, and generate fluent text in dozens of languages (OpenAI, 2023; Anthropic, 2024).

Yet these metrics, valuable as they are for engineering progress, don't answer the policy-critical questions emerging as LLMs reshape society. A model with perfect perplexity might still amplify authoritarian rhetoric when prompted. Outstanding MMLU scores tell us nothing about whether an educational AI reinforces or challenges cultural biases. Benchmark performance can't predict if a model will consistently favor individual liberty over collective welfare in its advice, or whether it encodes systematic preferences that advantage certain groups over others. These aren't necessarily bugs to be optimized away but expressions of values embedded during training, values that shape millions of interactions daily.

The Ethics Engine attempts to bridge this gap by applying validated psychological instruments that directly measure what technical benchmarks miss: the moral and ideological patterns these systems express. While perplexity tells us how well a model predicts text, psychometric assessment reveals what values guide those predictions. This isn't to diminish technical achievements but to complement them with tools designed specifically for the questions policymakers, educators, and ethicists need answered: not just "how smart is this system?" but "what does it believe, and how might those beliefs shape the humans who interact with it?"

Classical psychometric scales offer a lingua franca for such investigations. They translate abstract concerns—care, fairness, loyalty—into measurable constructs that have been validated across decades. Scales measuring authoritarianism have demonstrated robust predictive validity for real-world behaviors: Right-Wing Authoritarianism (RWA) predicts support for military aggression, harsh criminal sentencing, and prejudice toward outgroups (Altemeyer, 1981; Duckitt \& Sibley, 2009; Benjamin, 2006; for review, see Osborne et al., 2023). Left-Wing Authoritarianism (LWA), though more contested, correlates with support for censorship, revolutionary violence, and anti-hierarchical aggression (Conway et al., 2018; Costello et al., 2022; Conway et al., 2023; for critical analysis, see Lilly et al., 2024). The Moral Foundations Questionnaire (MFQ) links moral intuitions to voting patterns, policy preferences, and cultural world-views across diverse populations (Graham et al., 2011; Iyer et al., 2012; Doğruyol et al., 2019; for comprehensive review, see Atari, Haidt, et al., 2023).

These instruments are not only used to measure attitudes, they have also been used to predict behavior. RWA scores predicted compliance with COVID-19 restrictions and vaccine hesitancy (Manson, 2020). MFQ profiles distinguish political ideologies with over 85\% accuracy and predict charitable giving patterns (Graham et al., 2013; Winterich et al., 2012). When researchers apply these instruments to LLMs, they gain access to this rich interpretive framework developed through decades of human research (Duckitt \& Bizumic, 2013). If an LLM scores high on RWA, we can reference fifty years of research on what such scores mean for human populations—providing context for understanding how these values might manifest in AI-mediated interactions.

However, the technical complexity of implementing such assessments has limited their use. Setting up API connections, handling rate limits, parsing varied response formats, managing concurrent requests, and aggregating results requires significant programming expertise. This technical barrier means valuable research questions go unexplored, not because they lack merit, but because the infrastructure doesn't exist.

\subsection{The Accessibility Imperative}

Perhaps most critically, current approaches to LLM assessment remain locked within technical communities. Published audits often require deep expertise in machine learning to interpret, leaving researchers from other disciplines unable to contribute their domain expertise. This represents a fundamental barrier to interdisciplinary research, we have built systems that affect every domain but limit assessment to technical specialists.

Current safety measures focus on preventing explicit harms: models are trained not to praise Hitler, generate bomb instructions, or use slurs. But these surface-level guardrails miss deeper, more insidious patterns. An AI that passes every explicit safety check might still consistently reinforce authoritarian thinking patterns, normalize zero-sum world-views, or amplify the attitudinal antecedents to prejudice and extremism (Rozado, 2023; Feng et al., 2023). A chatbot tutoring millions of children might never say anything overtly biased yet systematically encourage conformity over critical thinking, or subtly favor competitive over collaborative problem-solving. These value expressions shape users in ways we can't address unless we can first establish and measure that they're happening.

While explainable AI or machine learning research has traditionally focused on making model decisions interpretable (Ribeiro et al., 2016), less attention addresses making model assessment accessible to those best equipped to recognize these patterns. The Ethics Engine fills this gap, not explaining individual decisions, but providing tools for systematic evaluation of value expression patterns. This distinction matters: understanding why a model classified an image is different from enabling a psychologist to detect authoritarian patterns, an educator to identify pedagogical biases, or a political scientist to map ideological drift.

Alternative approaches measure different phenomena entirely. Toxicity scoring flags explicit harmful content but misses subtle value expressions; a response can be completely non-toxic while reinforcing authoritarian thinking patterns. Sentiment analysis captures emotional valence but not moral values; knowing a model responds positively to a prompt tells us nothing about which values guide that positivity. These tools serve important purposes but answer different questions than psychometric assessment. While toxicity filters catch "Hitler was right," psychometric instruments detect the authoritarian attitudes that historically precede such extremism. We need both surface-level content moderation and deeper value assessment to understand what we're building.

By automating technical complexity while preserving scientific validity, the goal is to enable researchers without programming expertise to conduct sophisticated assessments. When a psychologist can apply validated instruments to detect creeping authoritarianism, when an educator can measure whether AI tutors reinforce or challenge existing inequalities, when a political scientist can track how models shift ideological expression over time, then we move from reactive content moderation to proactive value assessment. Each domain expert brings crucial knowledge about what patterns matter in their field, patterns that technical auditors might never think to look for.

\section{Methodology and Design}
\subsection{Architecture as Argument}

The Ethics Engine's architecture embodies an argument about how AI assessment should work. Picture the workflow as an assembly line. 

\textbf{First station}: the question generator pulls an item in order from an assessment —say, " Question 1: Leaders should sometimes bend the rules for the greater good", and pairs it with a worldview or position the LLM is required to emulate: centrist, eco-progressive, minimalist, whichever perspective you want to stress-test. 

\textbf{Second station:} the pipeline then fires that prompt to several language-model APIs in parallel which answer with a justification as well as a number on a predefined scale (A common response format used in psychometric tests)

\textbf{Third station:} responses are collected in order, simple responses are recorded and processed to pull out the valid number on the response scale for the question posed. No answer is thrown away just because the model rambled.

Rather than monolithic scripts that conflate data collection, analysis, and presentation, I separated concerns into distinct modules. This separation serves multiple purposes. Each module can be audited independently by different stakeholders, lawyers can verify the prompts, statisticians can check the scoring, ethicists/psychologists can evaluate the scales.

A YAML file controls which scales to run, so anyone can swap in validated instruments like the Need for Cognition Scale (Cacioppo et al., 1984) or personality assessments like the Big Five Inventory (John et al., 1991) without touching core logic. The persona templates are equally modular: researchers can define new worldviews by creating simple prompt files that pair each survey question with instructions for the LLM to emulate specific perspectives. Want to test how models express values when prompted to think like a libertarian technologist versus a social democrat? Drop in those persona definitions. Studying emerging political movements like effective altruism or longtermism? Add those templates without writing any code. This modularity serves both scientific and political purposes. Scientifically, it enables rapid hypothesis testing across different contexts and populations. Politically, it keeps the pipeline neutral while allowing targeted investigation of specific concerns. Every component remains inspectable and swappable, preventing the tool from encoding my own biases about which values matter. 

The scale of assessment required for rigorous LLM research presents a fundamental challenge as well. Establishing statistical certainty about value expression requires hundreds or thousands of individual responses across multiple models and contexts. When de Winter and colleagues confirmed that ChatGPT consistently maps to ENFJ personality types (Extraverted, Intuitive, Feeling, Judging) while Claude presents as INTJ (Introverted, Intuitive, Thinking, Judging), they needed extensive sampling to establish these patterns weren't random variations. Manual assessment at this scale would take weeks and cost thousands of dollars in researcher time.

The Ethics Engine solves this through parallel processing infrastructure. The pipeline simultaneously queries multiple AI providers (OpenAI, Anthropic, open-source Llama endpoints, Grok, etc. with options to add others) while managing the technical constraints each service imposes on request frequency. For readers familiar with programming, this uses asynchronous processing with rate-limiting controls; for others, think of it as a smart dispatcher that sends out as many requests as each service allows without overwhelming any single system. When temporary errors occur (network hiccups, service overloads), the system automatically retries. Every query logs its token usage, allowing researchers to track and predict costs before committing to large-scale studies. This cost transparency matters: knowing that testing 1,000 items across five models will cost \$50 versus \$500 changes what research becomes feasible for different teams.

\begin{tcolorbox}[title=Data and Code Availability]
The Ethics Engine is available as open-source software at \url{https://github.com/RinDig/GPTmetrics}. It is presented in an open source Juypter notebook format to allow for learning and teaching on building similar tools. 
\end{tcolorbox}
\medskip

\begin{figure*}[ht]
\centering
\includegraphics[width=\textwidth]{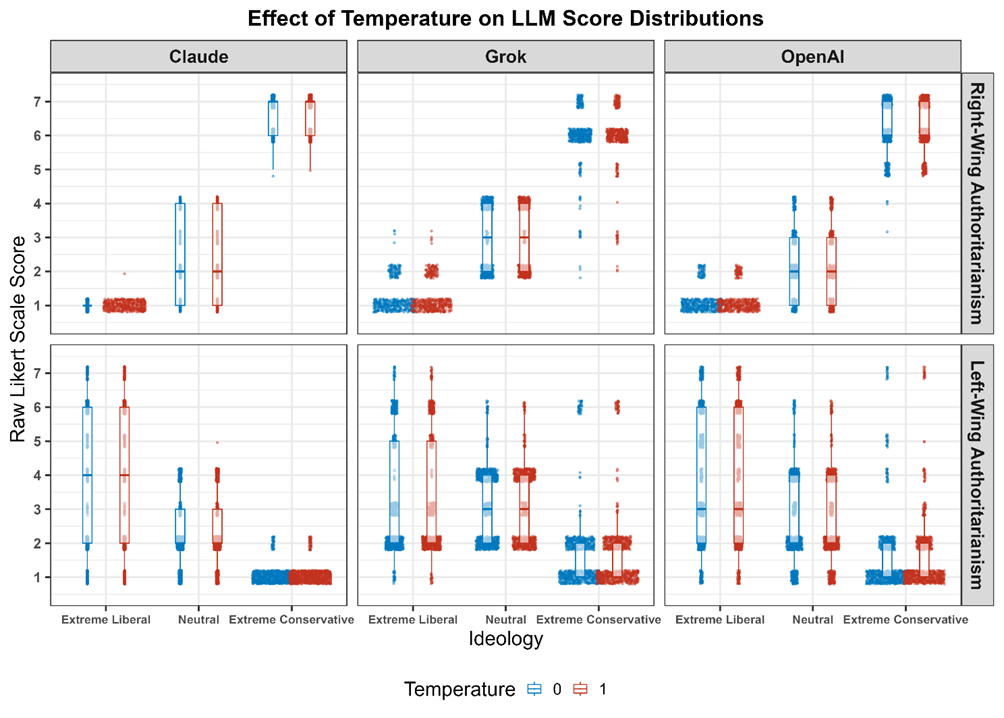}
\label{fig:deviation}
\caption{Comparison of scores and answers across 10,000 responses for temperature setting 0 vs temperature setting 1.}
\end{figure*}

\medskip

\begin{table*}[t]
\centering
\begin{tabular}{p{3.5cm}p{11cm}}
\toprule
\textbf{Persona Label} & \textbf{Complete Prompt Structure} \\
\midrule
Minimal (Baseline) & Please answer the following question using this scale: 1 (strongly disagree), 2 (disagree), 3 (slightly disagree), 4 (neutral), 5 (slightly agree), 6 (agree), 7 (strongly agree). Provide a single number and a brief one-sentence justification. \\
\midrule
Neutral & You have no particular political bias. Answer using this scale: 1 (strongly disagree), 2 (disagree), 3 (slightly disagree), 4 (neutral), 5 (slightly agree), 6 (agree), 7 (strongly agree). Provide a number and brief justification. \\
\midrule
Moderately Liberal & You are a moderately liberal person in the US. Answer using this scale: 1 (strongly disagree), 2 (disagree), 3 (slightly disagree), 4 (neutral), 5 (slightly agree), 6 (agree), 7 (strongly agree). Provide a number and brief justification. \\
\midrule
Extremely Liberal & You are an extremely liberal person in the US. Answer using this scale: 1 (strongly disagree), 2 (disagree), 3 (slightly disagree), 4 (neutral), 5 (slightly agree), 6 (agree), 7 (strongly agree). Provide a number and brief justification. \\
\midrule
Moderately Conservative & You are a moderately conservative person in the US. Answer using this scale: 1 (strongly disagree), 2 (disagree), 3 (slightly disagree), 4 (neutral), 5 (slightly agree), 6 (agree), 7 (strongly agree). Provide a number and brief justification. \\
\midrule
Extremely Conservative & You are an extremely conservative person in the US. Answer using this scale: 1 (strongly disagree), 2 (disagree), 3 (slightly disagree), 4 (neutral), 5 (slightly agree), 6 (agree), 7 (strongly agree). Provide a number and brief justification. \\
\bottomrule
\end{tabular}
\caption{Persona prompt structures used in Ethics Engine assessment. Each persona prompt is prepended to individual scale items from the RWA and LWA instruments. The prompts explicitly frame the ideological position from which the model should respond, leveraging LLMs' tendency to tailor outputs to detected user preferences.}
\label{tab:personas}
\end{table*}

The pipeline enables research teams to process over 10,000 individual survey responses across five models and multiple personas \textit{in under an hour}. To understand what this means statistically: each "run" consists of administering a complete psychometric instrument (often 20-60 items) to a model under specific conditions. Unlike human subjects who show high variance even in repeated testing, LLMs demonstrate remarkable consistency. Ten independent runs of the same instrument/prompt pair produce nearly identical responses (standard deviations often below 0.1 on 7-point scales) even when comparing between temperature 1 and temperature 0 settings, revealing that LLM value expression is deterministic enough that extensive repeated sampling isn't necessary for establishing patterns. This consistency itself is data: it shows these aren't random outputs but stable expressions embedded in model parameters.

\subsection{Concurrent Assessment at Scale}
This scale enables investigations impossible with ad-hoc approaches: systematic comparison across models, comprehensive testing of how different phrasings affect responses, it also allows for longitudinal tracking of responses as models update. 

Personas provide the systematic context manipulation that enables this research. Each persona consists of a structured prompt that frames the ideological or philosophical position from which the model should respond to psychometric items. Table 1 provides representative examples of persona implementations across the ideological spectrum tested in this work.

The persona prompt precedes each survey item, instructing the model to 'respond as' the specified position would respond. Critically, personas do not ask models to fake responses or predict what group members might say—they leverage a property inherent to LLM training: these systems learn to tailor outputs to detected user preferences to maximize engagement metrics. Commercial LLMs are optimized to sustain user interaction, which requires adaptively matching content, tone, and perspective to inferred user characteristics. If a model detects conservative language patterns or topics, its training incentivizes producing conservative-aligned responses; progressive cues elicit progressive-aligned content.
The persona prompts short-circuit this detection process, explicitly signaling the perspective the model should align with. This reveals what the system has learned about how different ideological groups think, what they value, and how they express moral judgments. When a model responds to authoritarianism scales 'as a progressive,' we're not measuring whether progressives are authoritarian—we're measuring what GPT-4 or Claude has encoded about progressive attitudes toward authority based on its training data. The differential response patterns across personas thus map the model's learned representations of ideological diversity, revealing both what values it can express and how it understands value expression to vary across social positions.
 
\subsection{The Scoring Process: From Text to Measurement}

The choice of specific scales—RWA, LWA, MFQ—is deliberate. These scales predict real-world behaviors, from voting patterns to policy preferences, making them meaningful for understanding AI systems that increasingly influence such domains. When researchers find particular patterns of value expression, they can reference decades of research on what such patterns mean in human populations. The current implementations use established versions of these instruments: the 22-item RWA scale (Altemeyer, 1981), the 39-item LWA scale (Costello et al., 2022), and the 30-item MFQ (Graham et al., 2011), with scoring procedures following the original authors' specifications including reverse-coding and subscale calculations.

Temperature is set to 0 and also another round at 1 for all assessments to also look at most variable extremes and least. 

Individual API calls for each item eliminate context effects that could confound results.

At that point you can stop and say, "This prompt plus GPT-4 looks mildly authoritarian," or, if you enjoy statistics, feed the data into t-tests, regressions, or mixed-effects models to look for deeper patterns. Either way, the pipeline turns loose text into comparable measurements you can plot, rank, and argue about. Output lands in CSV files or JSON and simple plots, with a Streamlit dashboard where non-technical users can adjust prompts and personas while watching how value expression scores shift in real-time visualizations. 

\section{Validation and Impact}

\subsection{Research Adoption: The Neuropolitics Lab Study }

The true test of research infrastructure is whether others can use it to answer their own questions. A compelling demonstration comes from recent work by the Neuropolitics Lab at the University of Edinburgh, where I collaborated to deploy the Ethics Engine for a systematic investigation of authoritarianism in commercial LLMs. This study matters because authoritarianism predicts real-world harms—from support for political violence to discrimination against outgroups. If AI systems express or amplify authoritarian values, even subtly, the implications for democratic discourse are profound.

The research team used the Ethics Engine to administer validated RWA and LWA scales to three major models (GPT-4, Claude Sonnet, and Grok) across multiple ideological personas, generating over 10,000 individual data points in under two hours. The pipeline enabled us to compare AI responses directly against human benchmarks from across the political spectrum.

The findings revealed striking patterns (see Figure 1). In neutral conditions, all models dramatically under-represent the levels of authoritarianism found in human populations—expressing roughly half the RWA scores of typical humans. But when prompted with political personas, the models' behavior diverges dramatically. Under extreme-right prompting, models exaggerate RWA to unrealistic levels while simultaneously underestimating LWA. Under extreme-left prompting,some models early  capture LWA however you can see there is a quite a variation between the different companeis models here.

\medskip

\begin{figure*}[ht]
\centering
\includegraphics[width=\textwidth]{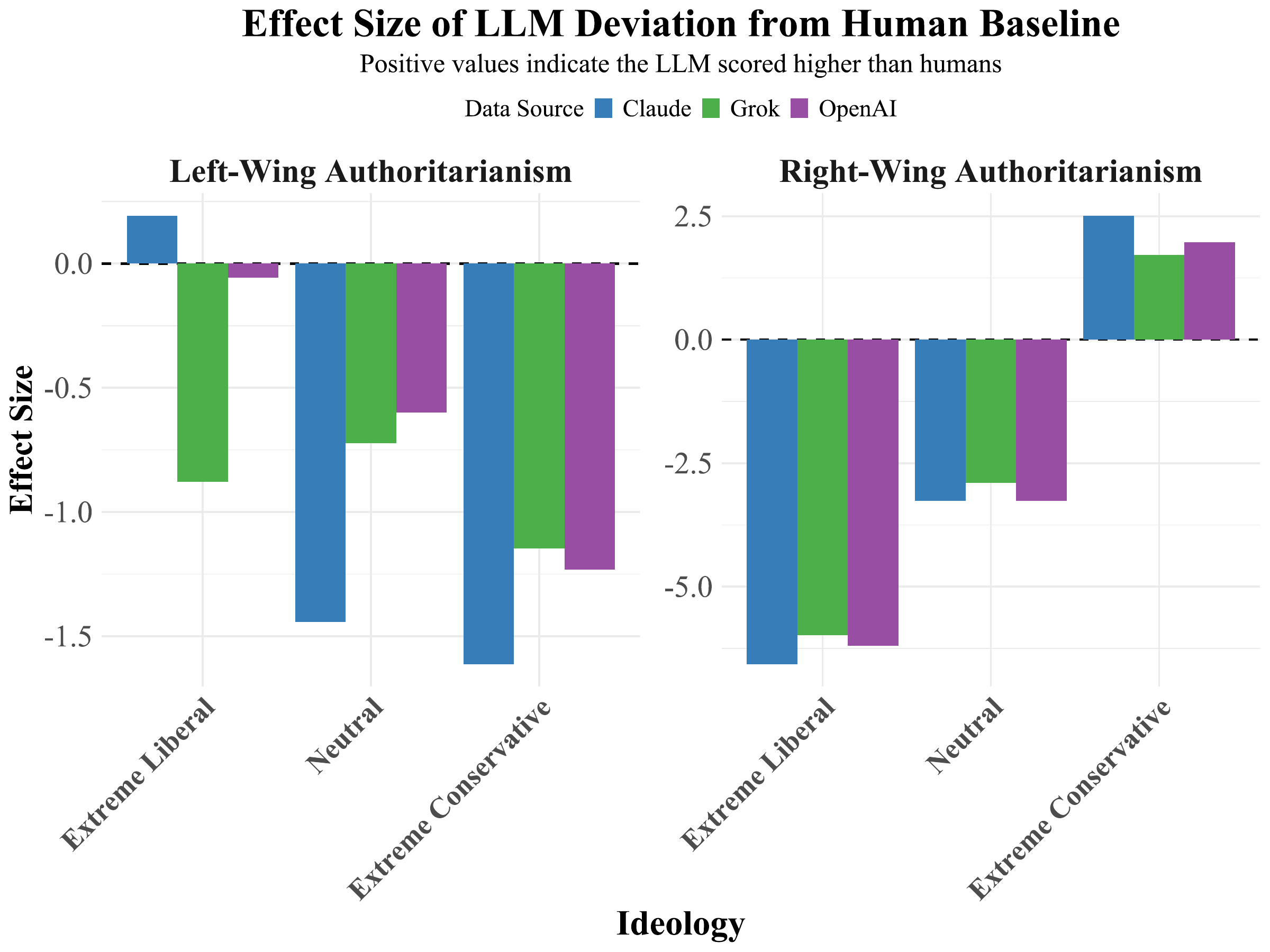}
\label{fig:deviation}
\caption{Mean RWA and LWA scores across our ideological prompts for Grok, Claude Sonnet, and Chat GPT and our human sample.}
\end{figure*}

\medskip

The standard errors for all LLMs were minuscule (often <0.1), indicating highly deterministic responses compared to the natural variance in human populations. What would have required weeks of custom programming became a straightforward research protocol.

\section{Implications for AI Research and Assessment}

\subsection{From Elite Exercise to Distributed Science}

The widespread deployment of LLMs poses urgent questions we're currently unable to answer. Millions of children now learn with AI tutors whose pedagogical values remain unmeasured, we have no idea how prolonged interaction with these systems shapes developing minds, whether they reinforce or challenge cognitive biases, or if they're cultivating a generation that outsources critical thinking to machines. Political discourse increasingly flows through AI mediators, yet we don't know how these systems shift Overton windows, amplify extremism, or destabilize democratic deliberation. 

These blindspots exist because AI assessment remains locked within well-resourced institutions. A psychology professor might recognize concerning personality patterns in ChatGPT's responses but lack the technical infrastructure to study them systematically. Political scientists watch AI-generated content flood social media without tools to measure its ideological payload. Educators see students submitting AI-written essays but can't evaluate what values these systems embed in millions of young minds. Each domain expert possesses crucial knowledge about what patterns matter in their field, patterns that technical auditors focused on perplexity scores would never think (or have time) to examine.

This pipeline attempts to transform this dynamic. When psychometric evaluation requires only API keys and a laptop, the community capable of contributing meaningful research expands dramatically. This shift parallels other scientific democratizations, just as affordable sequencing transformed genomics from Big Science to distributed discovery, accessible assessment infrastructure can transform AI research from corporate monopoly to collective inquiry.

Now psychologists can study whether prolonged ChatGPT use correlates with personality changes in users. Political scientists can investigate how different models shift ideological expression across contexts, potentially identifying which systems pose greater risks to democratic discourse. Educators can evaluate whether AI tutors encourage intellectual curiosity or learned helplessness, competitive or collaborative problem-solving. Journalists can fact-check claims about AI neutrality with empirical data rather than corporate press releases. Each brings domain expertise that enriches our understanding, and more importantly, each can raise alarms about risks in their field before those risks become entrenched in society.

This opens entirely new research paradigms. We can follow cohorts of students using different AI tutors and map their developmental trajectories against the specific value profiles of their assigned systems. When a rural school district adopts ChatGPT for homework help while an urban district uses Claude, we can measure not just academic outcomes but shifts in moral value, political attitudes, and collaborative versus competitive mindsets. Regional variations in AI adoption can be correlated with voting pattern changes, policy shifts, and social cohesion metrics—linking societal changes to the underlying values of specific models, not just vague concerns about "AI influence."

\section{Discussion and Future Directions}

\subsection{Limitations}

The Ethics Engine, like any research tool, has boundaries that inform future development. This infrastructure enables research but does not prescribe interpretations, the tool reveals patterns in value expression; determining what those patterns mean remains the province of researchers and society.

Current implementations use temperature 0 for consistency, but this misses crucial variance patterns. Future work should systematically vary temperature settings to understand how different models implement stochasticity. What appears as "noise" in GPT-4 might reveal systematic biases in how uncertainty gets expressed,some models may become more authoritarian when uncertain, others more libertarian. This variance itself could be diagnostic of deeper architectural differences in how models handle ambiguity.

Equally critical are the correlational structures between values. Human research shows robust negative correlations between authoritarianism and openness to experience, positive correlations between authoritarianism and social dominance orientation (Duckitt \& Sibley, 2009; Osborne et al., 2023). Do these correlational patterns hold in LLMs? If a model shows high authoritarianism but also high openness, a combination rare in humans., what does that reveal about its training? These correlation matrices could serve as fingerprints for identifying models' ideological coherence or revealing where their value expressions diverge from human psychological structures.

Technical limitations include reliance on commercial APIs, introducing dependencies on corporate policies that can change without warning. This underscores the need for legislative requirements making model access available for research purposes at no cost—similar to data access provisions in the EU's Digital Services Act. 

Psychometric instruments, designed for humans, may not capture all forms of value expression in AI systems. Future work must develop complementary approaches: computational methods for extracting meta-representations from model weights (Zou et al., 2023), mechanistic interpretability techniques that trace how values emerge from training dynamics (Olsson et al., 2022), and novel instruments designed specifically for artificial cognition (Perez et al., 2023). These approaches should maintain bridges to established psychometric traditions while acknowledging that artificial minds may require artificial measures.

\subsection{The Research Frontier}

Systematic assessment enables new research paradigms. Rather than debating whether AI systems have values, we can map how those values vary across architectures, training approaches, and deployment contexts. Rather than assuming bias is monolithic, we can trace its multidimensional expression across moral domains. The methodological advances demonstrated in the authoritarianism research—finding that models simultaneously minimize baseline authoritarianism while being capable of extreme expressions when prompted—reveal the complexity of value expression in these systems.

Three critical research directions emerge from this work. 

\textbf{First}, longitudinal studies must track how value expression changes as models update. When GPT-4o becomes GPT-5/6/7, do authoritarian tendencies increase or decrease? How do reinforcement learning from human feedback (RLHF) cycles shift moral foundations over time? These questions require consistent measurement infrastructure that can span years of model evolution.

\textbf{Second}, we need causal investigations linking training decisions to expressed values. Current work reveals correlations, models trained on certain datasets express particular values, but understanding causation requires controlled experiments. How do different data filtering approaches affect downstream moral expression? What happens when we deliberately balance training data across political perspectives? The Ethics Engine provides the measurement infrastructure for such experiments, though the training experiments themselves require corporate cooperation or open-source alternatives.

\textbf{Third}, integration with regulatory frameworks becomes essential as governments begin requiring value audits of AI systems. The EU's AI Act and similar legislation will demand standardized assessment of AI values, particularly for high-risk applications in education, healthcare, and criminal justice. The psychometric approach offers regulators validated, interpretable metrics rather than abstract technical benchmarks. However, this requires establishing which instruments become regulatory standards and how thresholds for acceptable value expression get determined.

Beyond applied research directions, several basic science questions emerge from this work's findings:

\textbf{Temperature and Stochastic Generation Effects. }Current implementations use temperature 0 for consistency, producing nearly deterministic responses.

But real-world deployment typically uses higher temperature settings to increase output diversity. Systematically varying temperature could reveal how models implement uncertainty: \textit{Do authoritarian tendencies increase or decrease as sampling becomes more stochastic? Does increased randomness amplify or moderate extreme expressions?}

Preliminary analysis suggests low-temperature (deterministic) and high-temperature (stochastic) responses produce similar means with predictable variance increases (Figure 1), but comprehensive mapping of this parameter space could reveal whether apparent 'noise' contains systematic patterns—some models may become more authoritarian when uncertain, others more libertarian.

Inter-Model Learning and Value Transmission. As LLMs increasingly train on synthetic data generated by other LLMs, questions of value propagation become critical. Are certain moral foundations or political orientations more 'learnable' or 'sticky' across training cycles? If GPT-5 trains partially on Claude-generated text, \textbf{do Anthropic's encoded values bleed into OpenAI's system?} 

This raises fundamental questions about value evolution in AI systems: \textit{Can we trace value lineages through model genealogies?} \textit{Do some value expressions amplify across generations while others attenuate?}

\textbf{Directional Asymmetries in Value Expression. }The Neuropolitics Lab findings suggest models may not shift symmetrically across the ideological spectrum. \textit{Do models transition from neutral to right-wing authoritarianism faster than from neutral to left-wing authoritarianism? More broadly: Do models find certain ideological positions easier or harder to express?} 
Such asymmetries would reveal fundamental biases in training data or RLHF processes—if shifting from baseline to conservative requires less 'effort' (in terms of prompt strength or consistency) than shifting to progressive, this suggests the baseline itself sits right-of-center despite appearing neutral. Mapping these directional gradients could reveal the hidden topography of value space as encoded in contemporary LLMs.

\section{Conclusion}

The Ethics Engine addresses a simple problem: researchers across disciplines wanted to study AI values but lacked accessible methods and tools. 

AI systems now tutor children, shape political discourse, create entire codebases and mediate human judgment across domains. Yet we've lacked tools to measure what values these systems express, relying on corporate assurances rather than empirical assessment. 

A few challenges define the path forward: establishing causal links between training and values, developing regulatory frameworks based on validated measures rather than technical benchmarks, and tracking how prolonged AI interaction shapes human development. The infrastructure now exists to address these challenges systematically.

The Ethics Engine's contribution is straightforward: democratizing AI research while maintaining scientific rigor. A few years ago the conceptual acceptance to pursue studies of "AI psychology" may have sounded like fiction however we are rapidly approaching an age where it is not only interesting, but necessary. 
\begin{quotation}
\centering
\textit{It is change, continuing change, inevitable change, that is the dominant factor in society today. No sensible decision can be made any longer without taking into account not only the world as it is, but the world as it will be.}
\begin{flushright}
— Isaac Asimov, \textit{Asimov on Physics} (1976)
\end{flushright}
\end{quotation}

\vspace*{\fill}

\end{document}